\documentclass[aps,showpacs]{revtex4}
\usepackage{graphics}
\usepackage{graphicx}
\usepackage{latexsym}
\usepackage{epsfig}
\begin{document}
\title{Testing Primordial Abundances With Sterile Neutrinos.}
\vskip2cm
\author{O. Civitarese and M. E. Mosquera}
\affiliation{{\small\it Department of Physics, University of La
Plata}  {\small\it c.c.~67 1900, La Plata, Argentina}}
\affiliation{{\small\it Faculty of Astronomy and Geophysics,
University of La Plata }{\small\it Paseo del Bosque s.n., La
Plata, Argentina}}
\date{\today}
\begin{abstract}
The mixing between sterile and active neutrinos is taken into
account in the calculation of Big Bang Nucleosynthesis. The
abundances of primordial elements, like D, $^3$He, $^4$He and
$^7$Li, are calculated by including sterile neutrinos, and by
using finite chemical potentials. It is found that the resulting
theoretical abundances are consistent with WMAP data on baryonic
densities, and with limits of LSND on mixing angles, only if
$^7$Li is excluded from the statistical analysis of theoretical
and experimental results.
\end{abstract}
 \maketitle key words: Sterile
Neutrinos, Primordial Abundances, WMAP data, LNSD, Big Bang
Nucleosynthesis.

\section{Introduction}

In recently published papers \cite{kishimoto,smith} the
sensitivity of the $^4$He primordial abundance, upon distortions
of the light neutrino spectrum induced by couplings with a sterile
neutrino, was analyzed. The effects due to the mixing between
sterile and active neutrinos \cite{kishimoto,smith} reflect upon
Big Bang Nucleosynthesis (BBN) in a noticeable manner. Previous
studies on this matter can be found in
\cite{cirelli1,cirelli2,cirelli3}, where the effects of mixing
upon BBN in presence of primordial leptonic asymmetry have been
investigated, and stringent limits on the mixing due to BBN have
been presented. Similar studies have been presented in
\cite{maxin1,maxin2}. A review on inclusion of sterile neutrino in
cosmology was presented in \cite{maxin3}.

The results of \cite{kishimoto} may be taken as a solid starting
point for a systematic analysis of the sterile-active neutrino
mixing upon cosmological observables, like the primordial
abundance of light elements. By the other hand, the mixing
mechanism between sterile and active neutrinos has been studied in
detail (see \cite{keranen03} and references therein), so that the
calculation of neutrino distribution functions can readily be
performed. The information about the neutrino distribution
function, in the flavor basis and at a given temperature, is an
essential element in the calculation of the neutron decay rate,
which is a critical quantity entering BBN \cite{BBF88,Esma91}.

Direct physical consequences upon BBN, due to the mixing between
active and sterile neutrinos, have been explored in
\cite{abazajian,bell}. Following the arguments presented in
\cite{abazajian}, and in the framework of the standard
cosmological model, sterile neutrinos would produce a faster
expansion rate for the Universe and a higher yield of $^4$He. This
is, indeed, a severe constraint on neutrino mixing since a higher
predicted abundance of $^4$He may be in conflict with
observational data \cite{abazajian}. Another constraint on
active-sterile neutrino mixing is the neutrino mass derived from
Cosmic Microwave Background Anisotropy (CMB) \cite{hannestad}.The
analysis of constraints presented in \cite{seljak} focus on the
mixing scheme at the level of the neutrino mass hierarchy, and it
suggests the adequacy of the non-degenerate mass hierarchy to set
limits on the mass difference between active and sterile
neutrinos, $\delta m^2_{a-s}$. The study of \cite{seljak} confirms
the notion about the convenience of the three active + one sterile
neutrinos scheme.

In standard BBN calculations, the mixing of sterile and active
neutrinos affects the leptonic fractional occupancies, which are
essential quantities appearing in the expression of the weak decay
rates. Thus, one needs to know, as input of the calculations, the
parameters of the proposed mixing scheme, the neutrino mass
hierarchy and the leptonic densities \cite{bell}. With these
elements one can calculate neutron-decay-rates and neutron
abundances, by assuming the freeze-out of weak interactions
\cite{BBF88,abazajian}. The effective number of neutrino
generations, $N_{\nu}$, is fixed by the analysis of CMB
\cite{steigman,barger1,barger2}. Current limits on the neutrino
degeneracy parameter, for light (electron) neutrinos, $\eta_l$
\cite{smith}, runs from $-0.1$ to $0.3$ \cite{barger1,barger2}.
For a detailed presentation of the formalism, in the context of
relic-neutrino asymmetry evolution see \cite{bell}.

In this work we focus on the calculation of the abundances of D,
$^3$He, $^4$He and $^7$Li, in presence of sterile-active neutrino
mixing in the three flavor scenario, and for the normal and
inverse neutrino mass hierarchies \cite{keranen03}. We have
compared the calculated values with data
\cite{muchosd,muchosh,muchosl} and determined the compatibility
between them by performing a $\chi^2$ statistical analysis. Since
the theoretical expressions depend on the mixing angle $\sin^2
2\phi$, the square mass difference $\delta m^2_{14}$ (normal mass
hierarchy) or $\delta m^2_{34}$ (inverse mass hierarchy), and the
baryonic density $\Omega_B h^2$, we have adopted the LSND limits
on the mixing angle \cite{eitel00,athana95,McGregor03} and the
WMAP results on the baryonic density \cite{wmap06}, as
constraints.

The paper is organized as follows. In Section \ref{formalism} we
briefly present the essentials of the formalism. Section
\ref{decay} is devoted to the calculation of the neutron decay
rate and BBN abundances. In Section \ref{results} we present and
discuss the results of the calculations. Conclusions are drawn in
Section \ref{conclusions}.

\section{Formalism}
\label{formalism}

The mixing between active neutrino mass eigenstates $\nu_i$
($i=1,2,3$), leading to neutrinos of a given flavor $\nu_k$
($k=\rm{light, medium, heavy}$), is described by the mixing matrix U
\cite{mixing}
\begin{eqnarray}
U&=&\left(
\begin{array}{ccc}
c_{13} c_{12}  & s_{12} c_{13}& s_{13} \\
-s_{12} c_{23}- s_{23} s_{13} c_{12} & c_{23} c_{12}-s_{23}
s_{13} s_{12}&s_{23}c_{13}\\
s_{23} s_{12}- s_{13} c_{23} c_{12}& -s_{23} c_{12}- s_{13} s_{12}
c_{23}& c_{23}c_{13}
\end{array}
\right), \label{U}
\end{eqnarray}
where $c_{ij}$ and $s_{ij}$ stand for $\cos \theta_{ij}$ and $\sin
\theta_{ij}$, respectively, and CP conservation is assumed
\cite{mixing}. To this mixing we add the mixing of a sterile
neutrino with: a) the neutrino mass eigenstate of lowest mass in
the normal mass hierarchy, $\nu_1$, and b) to the one of the
inverse mass hierarchy, $\nu_3$, by defining the mixing angle
$\phi$, such that the new mixing matrix $U$ is redefined as
$U(\phi)$ \cite{keranen03}
\begin{eqnarray}
U(\phi)&=& \left(
\begin{array}{cccc}
 c_{13} c_{12}  \cos \phi& s_{12} c_{13}& s_{13} & c_{13} c_{12}\sin \phi\\
\left(-s_{12} c_{23}- s_{23} s_{13} c_{12} \right)\cos \phi  &
c_{23} c_{12}-s_{23}
s_{13} s_{12}&s_{23}c_{13}&\left(-s_{12} c_{23}- s_{23} s_{13} c_{12} \right)\sin \phi \\
  \left(s_{23} s_{12}- s_{13} c_{23} c_{12}\right)\cos \phi &
-s_{23} c_{12}- s_{13} s_{12} c_{23} & c_{23}c_{13}& \left(s_{23}
s_{12}- s_{13}
c_{23} c_{12}\right)\sin \phi \\
-\sin \phi&0&0&\cos \phi
\end{array}
\right), \label{uphi}
\end{eqnarray}
for the normal mass hierarchy, and,
\begin{eqnarray}
U(\phi)&=&\left(
\begin{array}{cccc}
 c_{13} c_{12}  & s_{12} c_{13}&  s_{13}\cos \phi & s_{13}\sin \phi\\
-s_{12} c_{23}- s_{23} s_{13} c_{12} & c_{23} c_{12}-s_{23}s_{13}
s_{12}&  s_{23}c_{13}\cos \phi& s_{23}c_{13}\sin \phi\\
s_{23} s_{12}- s_{13} c_{23} c_{12} & -s_{23} c_{12}- s_{13}
s_{12}
c_{23} &  c_{23}c_{13}\cos \phi&  c_{23}c_{13}\sin \phi\\
0&0&-\sin \phi&\cos \phi
\end{array}
\right),  \label{uphi-i}
\end{eqnarray}
for the inverse mass hierarchy. The mixing between neutrino mass
eigenstates, and particularly the inclusion of the sterile
neutrino as a partner of the light neutrino, affects the
statistical occupation factors of neutrinos of a given flavor. The
equation which determines the structure of the neutrino occupation
factors, in the basis of mass eigenstates and for an expanding
Universe, can be written \cite{kirilova88}:
\begin{eqnarray}
\left(\frac{\partial f}{\partial t}- {\rm H} E_\nu \frac{\partial f}
{\partial E_\nu}\right)&=& \imath \left[H_0,f\right], \label{equf}
\end{eqnarray}
where  $t$ is time, ${\rm H}$ is the expansion rate of the
Universe, defined as ${\rm H}= \sqrt{\frac{ 4 \pi^3 N}{45
M_{\rm{Planck}}^2}}  T^2= \mu_P T^2$, $T$ is the temperature,
$E_\nu$ is the energy of the neutrino, and $H_0$ is the
unperturbed mass term of the neutrino's Hamiltonian in the rest
frame. The initial condition is fixed by defining the occupation
numbers at the temperature $T_0= 3$ MeV \cite{dolgov97},
\begin{eqnarray}
\label{cinorm} \left. \left(
\begin{array}{cccc}
f_{11}&f_{12}&f_{13}&f_{14}\\
f_{21}&f_{22}&f_{23}&f_{24}\\
f_{31}&f_{32}&f_{33}&f_{34}\\
f_{41}&f_{42}&f_{43}&f_{44}
\end{array}
\right)\right|_{T_0} &=& \frac{1}{1+e^{E_\nu/T_0-\eta}} \left(
\begin{array}{cccc}
\cos^2\phi&0&0&\sin \phi\cos \phi\\
0&1&0&0\\
0&0&1&0\\
\sin \phi\cos \phi&0&0&\sin^2 \phi
\end{array}
\right),
\end{eqnarray}
for the normal mass hierarchy, and
\begin{eqnarray}
\label{ciinv} \left. \left(
\begin{array}{cccc}
f_{11}&f_{12}&f_{13}&f_{14}\\
f_{21}&f_{22}&f_{23}&f_{24}\\
f_{31}&f_{32}&f_{33}&f_{34}\\
f_{41}&f_{42}&f_{43}&f_{44}
\end{array}
\right)\right|_{T_0} &=& \frac{1}{1+e^{E_\nu/T_0-\eta}} \left(
\begin{array}{cccc}
1&0&0&0\\
0&1&0&0\\
0&0&\cos^2\phi&\sin \phi\cos \phi\\
0&0&\sin \phi\cos \phi&\sin^2 \phi
\end{array}
\right),
\end{eqnarray}
for the inverse mass hierarchy.

To obtain the solutions of Eq.(\ref{equf}) we have written the
source term, that is the commutator in the r.h.s of
Eq.(\ref{equf}), in terms of the square mass differences, $\delta
m_{ij}^2=m^2_i-m^2_j$:
\begin{eqnarray}
\left[H_0,f\right]&=& \frac{1}{2 p}\left(
\begin{array}{cccc}
0&\delta m_{12}^2 f_{12}&\delta m_{13}^2 f_{13}&\delta m_{14}^2 f_{14}\\
-\delta m_{12}^2 f_{21}&0&\delta m_{23}^2 f_{23}&\delta m_{24}^2 f_{24}\\
-\delta m_{13}^2 f_{31}&-\delta m_{23}^2 f_{32}&0&\delta m_{34}^2 f_{34}\\
-\delta m_{14}^2 f_{41}&-\delta m_{24}^2 f_{42}&-\delta m_{34}^2
f_{43} &0
\end{array}
\right).
\end{eqnarray}
The value of the mixing angle $\theta_{13}$ is constrained by the
upper limit given by \cite{exp} so that $\tan \theta_{13} \leq
10^{-3}$. The solution in the basis of mass eigenstates is
\begin{eqnarray}
f_{ii}&=&\frac{\rm{const}}{1+e^{E_\nu/T-\eta}}\nonumber \\
f_{ij}&=&\frac{\rm{const}}{1+e^{E_\nu/T-\eta}}\, {\rm
exp}\left[\imath \frac{\delta m^2_{ij}}{6\mu_P}
\frac{T}{E_\nu}\left(\frac{1}{T^3}-\frac{1}{T^3_0}\right)\right]
\end{eqnarray}
where the normalization constants are fixed by the initial
conditions ($T=T_0$). The formal solution for the occupation
number in the flavor basis, for the light neutrino flavor in the
normal mass hierarchy, is:
\begin{eqnarray}
f_{l} &=&\frac{1}{1+e^{E/T-\eta}} \left\{ 1+ \cos^2\theta_{13}
\cos^2\theta_{12}\frac{\sin^2
2\phi}{2}\left[\cos\left(\frac{\delta m_{14}^2}{6\mu_P}
\frac{T}{E}\left(\frac{1}{T^3}-\frac{1}{T_0^3}\right)\right)-1\right]\right\},\label{fln}
\end{eqnarray}
and
\begin{eqnarray}
f_{l}&=&\frac{1}{1+e^{E/T-\eta}} \left\{ 1+ \sin^2\theta_{13}
\frac{\sin^2 2\phi}{2}\left[\cos\left(\frac{\delta
m_{34}^2}{6\mu_P} \frac{T}{E}
\left(\frac{1}{T^3}-\frac{1}{T_0^3}\right)\right)
-1\right]\right\},\label{fli}
\end{eqnarray}
in the inverse mass hierarchy.

In the above expressions, $\eta$ is the ratio between the neutrino
chemical potential and the temperature. This parameter depends on
the adopted value of the leptonic number $\rm{L}$
\cite{smith,kishimoto}. Explicit expressions of $\eta$ versus
$\rm{L}$ can be found in \cite{smith}. In the present context we
have taken $\eta$ as an input for the calculations (see  section
\ref{decay}).

\section{Decay Rates and neutron abundance}
\label{decay}

In the following we shall outline the main steps of the calculation
of neutron decay rates, for the electroweak processes
$n+e^+\rightarrow p+\overline{\nu}$ and $n+\nu\rightarrow p+e^-$.
The starting point is the calculation of the reduced rates
$\lambda_{\pm}$\cite{BBF88}
\begin{eqnarray}
\lambda\left(n+\nu\rightarrow p+ e^- \right)=\lambda_- &=&
\lambda_0\int_0^\infty dp_\nu p_\nu E_\nu
p_e E_e \left(1-f_e\right) f_l, \\
\lambda\left(n+e^+\rightarrow p+ \overline{\nu} \right)=\lambda_+
&=& \lambda_0\int_0^\infty dp_e p_\nu E_\nu p_e E_e
\left(1-f_l\right) f_e,
\end{eqnarray}
and the total neutron to proton decay rate
\begin{eqnarray}
\lambda_{np}(y) &=&\lambda_- (y)+\lambda_+ (y)\nonumber \\
&=& 2 \lambda_0 \left[e^{\eta} \left( 1-\alpha\frac{\sin^2 2
\phi}{2}\right)+1 \right] \frac{\Delta
m_{np}^5}{y^3}\left(1+\frac{6}{y}+\frac{12}{y^2}\right)\nonumber \\
&&+\lambda_0\Delta m_{np}^5 \frac{\sin^2 2 \phi}{2} e^{\eta}
\int_0^\infty dq q^2 \left(q+1\right)^2 e^{-q y}
g\left(q,y\right), \label{lnp}
\end{eqnarray}
at lowest order in the quantity $e^{\eta}$. For the sake of
convenience we have introduced the more compact notation
$f_l=\left(1+e^{E_\nu/T-\eta}\right)^{-1} \left\{1-\alpha \frac{
\sin^2 2 \phi}{2} +\frac{\sin^2 2 \phi}{2}
g\left(E_\nu,T\right)\right\}$ with $ \alpha=c_{13}^2c_{12}^2$ for
the normal mass hierarchy and $\alpha=s_{13}^2$ for the inverse
mass hierarchy, respectively. The function $g(E_\nu,T)$ is the
factor which contains the temperature $T$ and the energy $E_\nu$
in Eqs.(\ref{fln}) and (\ref{fli}), and the variable $y$ is
defined as $y=\frac{\Delta m_{np}}{T}$. The details of the
calculations have been discussed elsewhere \cite{nosotros}, for
the case of two neutrino mass eigenstates. The final expression
for the neutron to proton decay rate is obtained by fixing the
normalization $\lambda_0$, of Eq.(\ref{lnp}), from the neutron
half-life
\begin{eqnarray}
\frac{1}{\tau}&=& \frac{4 \lambda_0 \Delta m_{np}^5}{255},
\end{eqnarray}
and the result is
\begin{eqnarray}
\lambda_{np}(y) &=& \frac{255}{2\tau}
\left[e^{\eta}\left(1-\alpha\frac{\sin^2 2 \phi}{2}\right)+1
\right] \left(\frac{1}{y^3}+\frac{6}{y^4}+\frac{12}{y^5}\right)
+\frac{255}{4 \tau} \frac{\sin^2 2 \phi}{2}e^{\eta} \int_0^\infty
dq q^2 \left(q+1\right)^2 e^{-q y}g\left(q,y\right), \label{lnp1}
\end{eqnarray}
in units of ${\rm{sec}}^{-1}$. Following Ref.\cite{BBF88}, the
neutron abundance, until the freeze-out of weak interactions, is
expressed in terms of the neutron to proton decay rate,
$\lambda_{np}$ of Eq.(\ref{lnp1}) as
\begin{eqnarray}
X_{\rm{neutrons}}=\int^\infty_0 dw \;
e^{w+\eta}\left(\frac{1}{1+e^{w+\eta}}\right)^2 e^{-\left(\mu_P
\Delta m_{np}^2 \right)^{-1}\int_{w}^\infty du u
\left(1+e^{-u-\eta}\right) \lambda_{np} (u)}.
\end{eqnarray}
The quantity $X_{\rm{neutrons}}$ is, therefore, a function of
$\lambda_{np}$ and, consequently, of the occupation factors $f_l$,
which contain the information about the mixing between active and
sterile neutrinos. The next step consists on the calculation of
primordial nuclear abundances. The method to calculate the BBN
abundances was presented in Ref \cite{Esma91}. It is a
semi-analytic approach based on the balance between production and
destruction of a given nuclear element, which requires the
knowledge of $X_{\rm{neutrons}}$. For details we refer the reader
to \cite{Esma91}.

The above presented framework shows that the calculation of
primordial abundances may indeed be taken as a tool to test leptonic
mechanisms, like the mixing between sterile and active neutrinos, as
it has been pointed out by Kishimoto et al. \cite{kishimoto}.

\section{Results and Discussion}
\label{results}

To perform the calculations we have adopted the oscillation
parameters determined from SNO, SK and CHOOZ measurements
\cite{exp}. The mixing with the sterile neutrino, represented by
the mixing angle $\phi$, is taken as an unknown variable, within
the limits fixed by the LSND data
\cite{eitel00,athana95,McGregor03}. The mass splitting $\delta
m^2_{14}$ (or $\delta m^2_{34}$) was taken from the analysis given
by Ker\"anen et al. \cite{keranen03}. The actual value is fixed at
$\delta m^2=10^{-11}$ eV$^2$. We have then calculated the neutron
abundance, by applying the formalism of the previous section. The
primordial abundances of D, $^3$He, $^4$He and $^7$Li, have been
calculated as described in \cite{Esma91}. The baryonic density
$\Omega_B h^2$ (see Ref.\cite{nosotros}) was varied within the
limits $0.010 < \Omega_B h^2 < 0.035$. Concerning the value of
$\eta$ we have varied it in the interval determined by the allowed
values of the potential lepton number, ${\cal{L}}=2
\rm{L}_{{\nu}_e}+\rm{L}_{{\nu}_{\mu}}+ \rm{L}_{{\nu}_{\tau}}$,
that is $0.0 \leq {\cal{L}} \leq 0.4 $ \cite{kishimoto,smith}. In
the present calculations we have adopted the values $0.0 \leq \eta
\leq 0.07$ which are consistent with the densities $0.0 \leq
\rm{L}_{{\nu}_e} \leq 0.05 $ \cite{smith}.

To determine the allowed values of the mixing angle $\phi$ we have
performed a $\chi^2$-minimization, after computing the primordial
abundances. The data have been taken from
Refs.\cite{muchosd,muchosh,muchosl}. The results are shown in
Figure 1, insets 1.(a) and 1.(b). The curves are the contour plots
for results with comparable values of $\chi^2$. Figure 1.(a) shows
the results obtained by the $\chi^2$-analysis of theoretical and
experimental values \cite{muchosd,muchosh,muchosl}, including data
on $^7$Li. Figure 1.(b) shows the results of the statistical
analysis performed with the exclusion of $^7$Li. In the first
case, Figure 1.(a), the absolute minimum is located at $\sin^2
2\phi=0.000 \pm 0.026$, and $\Omega_B h^2=0.0253 \pm 0.0015$, both
set of results have been obtained by using the solution
(\ref{fln}) for the occupations. The smallness of the mixing angle
does not contradict LSND results
\cite{eitel00,athana95,McGregor03}, but the value of the baryonic
density is outside the limits determined by  WMAP \cite{wmap06},
that is: ${(\Omega_B h^2)}_{\rm{WMAP}}=0.0223\pm 0.0008$. This
disagreement between theory and data may be caused by large
uncertainties in the $^7$Li-data. As pointed out by Richard et al.
\cite{richard05}, the validity of the data on $^7$Li may be
questioned by the uncertainties inherent to the physics of $^7$Li
in the interior of the stars, i.e; the turbulent transport in the
radiative zone of stars. In contrast, the situation improves if
the data on $^7$Li are removed at the time of performing the
statistical analysis (Figure 1.(b)). For this case, the best value
of the mixing angle is $\sin^2 2\phi=0.018 \pm 0.098$, and the
baryonic density corresponding to the minimum, ${\Omega_B
h^2}=0.0216\pm 0.0017$, is indeed consistent with the WMAP data.
The anomalous feature associated with the inclusion of $^7$Li in
the set of data persists if other elements are removed from the
data. We have verified it by systematically removing, one at the
time, the abundances of D, $^3$He, and $^4$He, and keeping the
data on $^7$Li. In all cases the location of the minimum lies
closer to the one of Figure 1.(a). For the case of inverse mass
hierarchy the occupation factor (\ref{fli}) is strongly
constrained by the value of $\theta_{13}$ and the difference with
respect to the thermal occupation factor vanishes. It means that
the contour plot shows, for the inverse mass hierarchy, parallel
lines to the vertical axis ($\sin^2 2\phi$), since all possible
values of $\sin^2 2\phi$ are allowed by Eq. (\ref{fli}) when
$\sin^2\theta_{13}\rightarrow 0$. For both the normal and inverse
hierarchy solutions, Eq. (\ref{fln}) and (\ref{fli}), particle
number conservation was enforced, on the average, by the factor
$\mu_P$ (see its definition following Eq.(\ref{equf})). Because of
the high temperature we have not included collision terms in Eq.
(\ref{equf}).

Similar results, related to the abundance of $^7$Li, have been
obtained in the calculations of nuclear abundances in the context
of cosmological models \cite{cuoco04,coc04,mosquera06}, and also
in the case of a two neutrino mixing \cite{nosotros}.

\begin{figure}[!h]
\begin{center}
\begin{tabular}{cc}
\epsfig{file=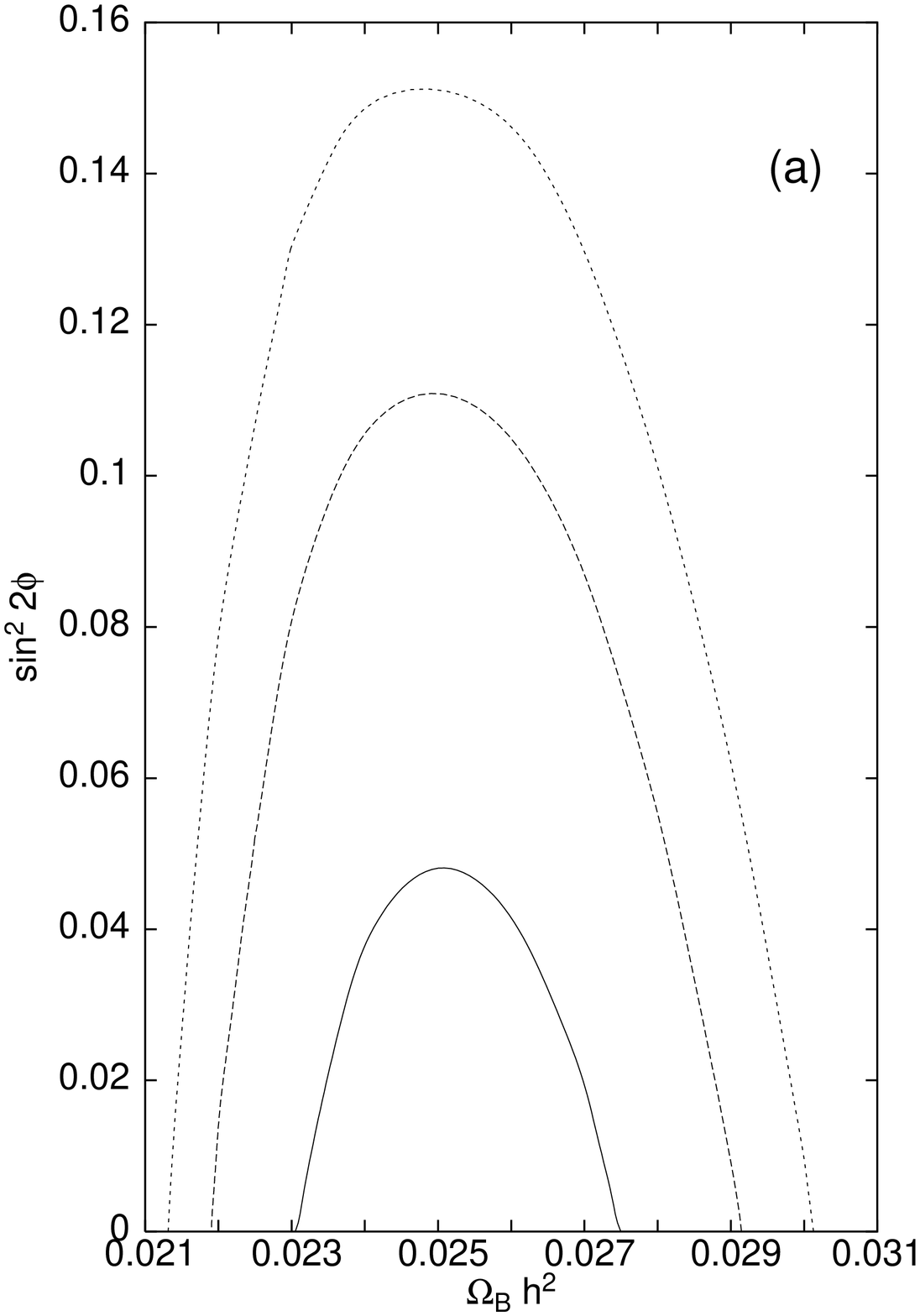, angle=0, width=250pt}&
\epsfig{file=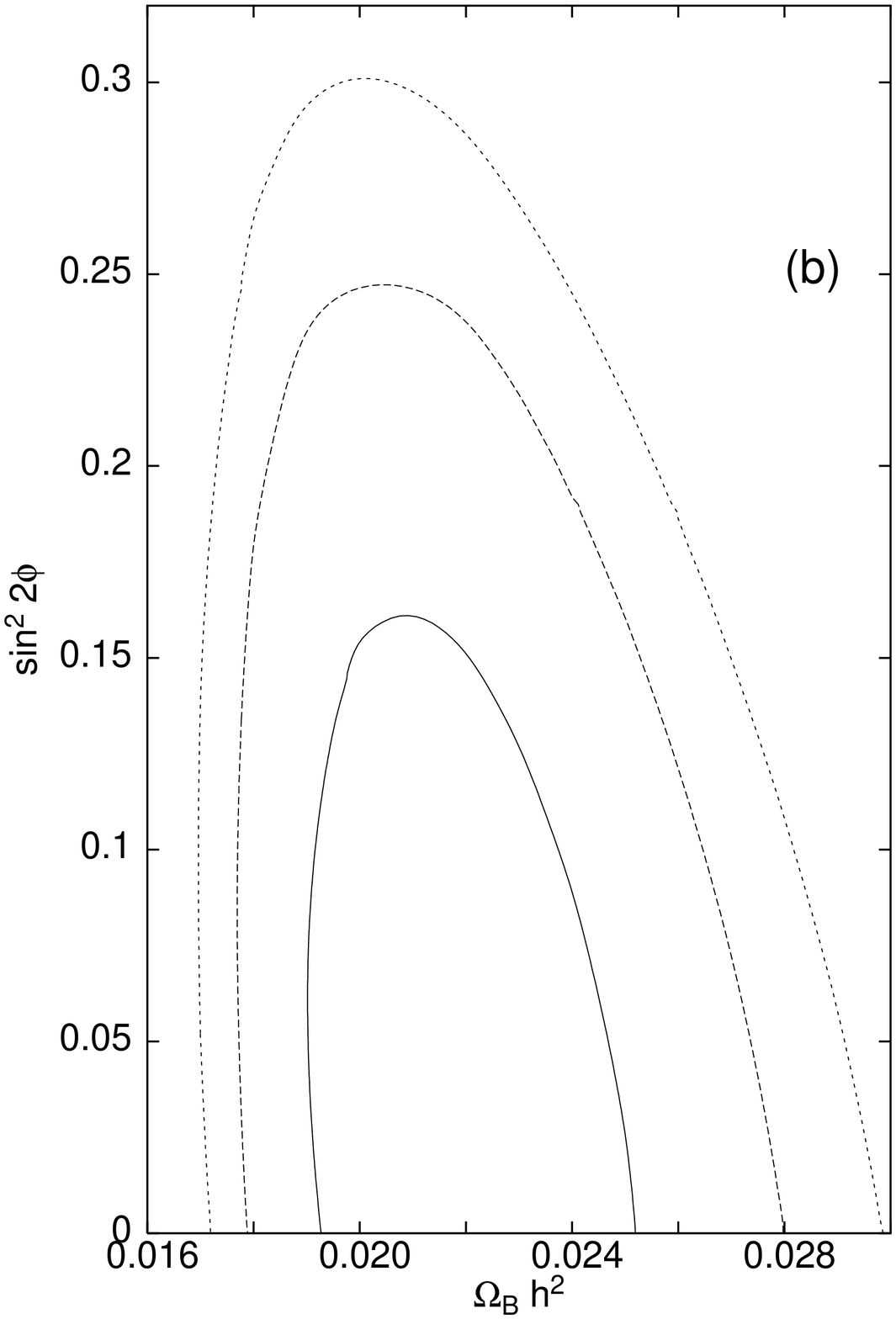, angle=0, width=250pt}
\end{tabular}
\caption{Statistical analysis of the calculated nuclear
abundances. The curves are the contour plots for results with
comparable $\chi^2$ values. The contours correspond to increasing
values of $\chi^2$, from bottom to top. The calculations were
performed by taking the sterile-active neutrino mixing, $\sin ^2
2\phi$, and the baryonic density, $\Omega_B h^2$, as variables.
The inset 1.(a) shows the results obtained by the $\chi^2$
analysis of theoretical and experimental values, including data on
$^7$Li. Inset 1.(b) shows the results of the statistical analysis
performed with the exclusion of $^7$Li. The results shown in the
figure have been obtained with the solution corresponding to the
normal mass hierarchy.} \label{figuras}
\end{center}
\end{figure}

\begin{figure}[!h]
\begin{center}
\epsfig{file=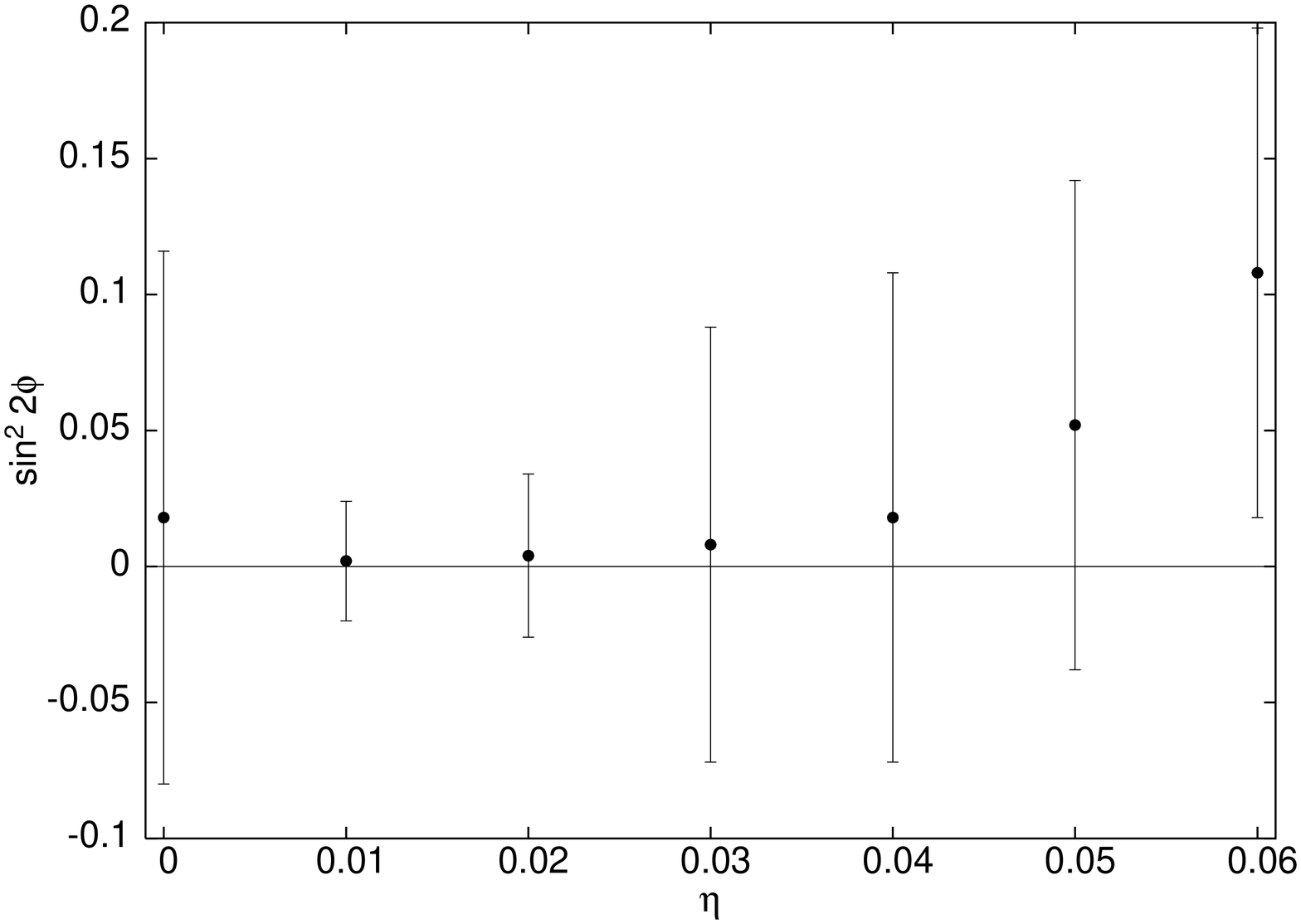, angle=-90, width=250pt} \caption{Best values
of the mixing angle, $\sin^2 2\phi$, determined from the $\chi^2$
analysis of the calculated abundances, as functions of the
parameter $\eta$} \label{figure2}
\end{center}
\end{figure}

To investigate the dependence of the above results on the
parameter $\eta$, we show in Figure \ref{figure2} the values of
the mixing angle $\sin^2 2 \phi$, obtained from the $\chi^2$
analysis, as functions of the chemical potential. The calculations
have been performed by excluding the data on $^7$Li. Our present
results are very much in agreement with the results reported in
Ref. \cite{BBF88}, since the values shown in Figure \ref{figure2}
display a small variation in a relatively large domain of values
of $\eta$.
Finally, the best values of the baryonic density and the mixing
angle, both with and without including $^7$Li in the analysis, are
shown in Table \ref{table} as functions of $\eta$. In agreement
with the expectations of \cite{BBF88}, and with our owns, both
sets of results do not differ much, or at least they do not show a
pronounced dependence, with respect to the chemical potential.
\begin{table}[!h]
\begin{center}
\begin{tabular}{c|c|c||c|c}
&\multicolumn{2}{c||}{All data}&\multicolumn{2}{c}{All data but
    $^7$Li}\\ \hline
$\eta$& $\Omega_B h^2$ & $\sin^2 2 \phi$& $\Omega_B h^2$ & $\sin^2
2\phi$ \\ \hline
$0.00$ & $0.0253 \pm 0.0015$ &$0.000\pm 0.026$
&$0.0216\pm 0.0017$&$0.018\pm 0.098$   \\ \hline $0.01$ & $0.0250
\pm 0.0014$ &$0.000\pm 0.010$ &$0.0216\pm 0.0020$&$0.002\pm 0.022$
\\ \hline $0.02$ & $0.0248 \pm 0.0014$ &$0.000\pm 0.015$
&$0.0218\pm 0.0020$&$0.004\pm 0.030$  \\ \hline
 $0.03$ & $0.0246 \pm 0.0012$
&$0.000\pm 0.034$ &$0.0216\pm 0.0018$&$0.008\pm 0.080$  \\ \hline
 $0.04$ & $0.0244 \pm 0.0016$
&$0.000\pm 0.039$ &$0.0216\pm 0.0019$&$0.018\pm 0.090$  \\ \hline
 $0.05$ & $0.0244 \pm 0.0016$
&$0.000\pm 0.056$ &$0.0216\pm 0.0017$&$0.052\pm 0.090$  \\ \hline
 $0.06$ & $0.0244 \pm 0.0016$
&$0.000\pm 0.101$ &$0.0216\pm 0.0018$&$0.108\pm 0.090$  \\ \hline
\end{tabular}
\caption{Best values of the mixing angle and of the baryonic
density, determined from the $\chi^2$ analysis of the calculated
abundances, as functions of the parameter $\eta$. Left and right
sides of the table show the results obtained with and without
considering the data on $^7$Li, respectively. } \label{table}
\end{center}
\end{table}

\section{Conclusions}
\label{conclusions}

In this work we have calculated BBN abundances, by including the
mixing between active and sterile neutrinos. As pointed out by
Kishimoto et al, the BBN abundances are sensitive to
active-sterile neutrino mixing, indeed. Kishimoto et al.
\cite{kishimoto} have demonstrated the sensitivity of the $^4$He
abundance on the distortion of the light neutrino spectrum
produced by the mixing with a sterile neutrino. In our case, the
statistical analysis of the compatibility between theoretical and
observed nuclear abundances, indicates the existence of a clear
sensitivity of the results upon active-sterile neutrino mixing,
too. In performing our analysis, we have considered the WMAP
baryonic density, together with the LSND constraint on the
sterile-active neutrino mixing. The comparison between calculated
and observed abundances indicates some sort of anomaly in the
abundance of $^7$Li. Similar difficulties, related to the
determination of the abundance of $^7$Li, have been reported
previously \cite{richard05}, in the context of the physics of the
interior of stars. We found that the consideration of the
abundance of $^7$Li, in presence of active-sterile neutrino
mixing, excludes the WMAP value of the baryonic density. This
exclusion is not observed when the other nuclear abundances are
not included in the analysis.

\section{Acknowledgements}

This work has been partially supported by the National Research
Council (CONICET) of Argentina. Discussions with Professor Jukka
Maalampi (University of Jyv\"askyl\"a) are gratefully
acknowledged.

\end{document}